\documentclass[12pt]{iopart}

\usepackage{iopams}  
\usepackage{graphicx}
\usepackage{graphics}
\usepackage{amssymb}
\usepackage{epsfig}
\usepackage{color}
\begin{document}

\title[On the capillary self-focusing in a microfluidic system]{On the capillary self-focusing in a microfluidic system}
\author{
M. Hein$^1$,
R. Seemann$^1$
\footnote{Corresponding author: r.seemann@physik.uni-saarland.de},
and S. Afkhami$^2$
\footnote{Corresponding author: shahriar.afkhami@njit.edu}
}
\address{$^1$ Experimental Physics, Saarland University, 66123 Saarbr\"{u}cken, Germany}
\address{$^2$ Department of Mathematical Sciences,
New Jersey Institute of Technology, Newark, NJ 07102, USA}

\begin{abstract}
A computational framework is developed to address
capillary self-focusing in Step Emulsification. 
The microfluidic system consists of a single shallow and wide microchannel
that merges into a deep reservoir. A continuum approach coupled with
a volume of fluid method is used  to model the capillary self-focusing effect.
The original governing equations are
reduced using the Hele--Shaw approximation.
We show that the interface between the two fluids takes
the shape of a neck narrowing in the flow direction just before 
entering the reservoir, in agreement with our experimental observations.
Our computational model relies on the assumption that the pressure at the boundary,
where the fluid exits into the reservoir, is the uniform pressure in the reservoir.
We investigate this hypothesis by comparing the numerical results
with experimental data. We conjecture that the pressure boundary condition 
becomes important when the width of the neck is comparable to the depth of the microchannel.
A correction to the exit pressure boundary condition is then proposed, which
is determined by comparison with experimental data.
We also present the experimental observations and the numerical results
of the transitions of breakup regimes.
\end{abstract}

\vspace{2pc}
\noindent{\it Keywords}: Microfluidics, Step Emulsification, Capillary Focusing, Hele--Shaw, 
Volume Of Fluid Simulation

\maketitle

\begin{figure}[t]
\centering
\includegraphics[width=105.mm]{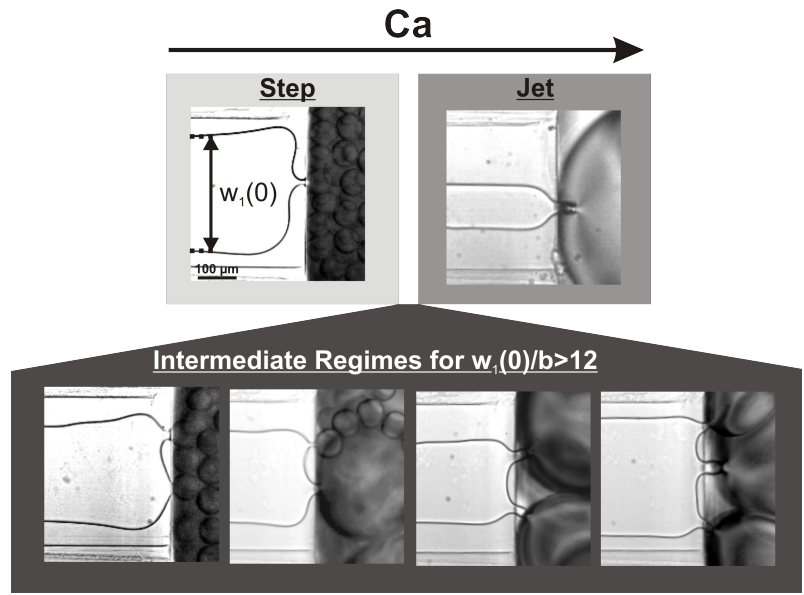}
\caption{Breakup regimes observed in step emulsification droplet generation. 
At low $w_1(0)/b$, with $b$ the depth of the shallow channel, the step and the 
jet emulsification regimes are observed for low and high capillary numbers, Ca, respectively. 
Insert: For $w_1(0)/b > 12$, the coexistence of multiple breakup sites on a single filament is observed.}
\label{fig:overview}         
\end{figure}
\section{Introduction: Background and the Mathematical Model}
Step emulsification is a microfluidic technique 
for generating monodisperse droplets with highly controllable
sizes and production rates (Priest \etal 2006). The process
relies on breakup of a long and wide confined sheet of fluid, 
surrounded by a second fluid, at a step geometry, consisting of a single shallow and wide microchannel
that abruptly merges into a deep reservoir (see Fig.~\ref{fig:overview}). At the step, 
the interface between the two fluids takes the shape of a tongue 
narrowing in the flow direction; we call this the capillary self-focusing. 
When the self-focused pointed tongue enters into
the reservoir, the interface becomes unstable resulting in the formation
of droplets at the step. 
This can be explained by the fact that in the focusing region, a cylindrical thread
is formed and becomes unstable by the resulting lack of confinement, as it enters the
reservoir, presumably by a Rayleigh-Plateau instability mechanism (Li \etal 2015).
Previously, we showed that
%For microfluidics applications, 
the width of this focusing region is directly related to the transition of
breakup regimes (Hein \etal 2015a), namely the jet- and step-breakup regimes,
where the breakup transitions from a quasi-steady (jet-breakup) to a transient (step-breakup) regime
(see Fig.~\ref{fig:overview} top panels).  
%This interfacial instability and the breakup mechanism are only poorly understood. 
Here, we aim to extend our investigation of this self-focusing phenomena
and our understanding by comparing the numerical results with 
experimental observations. 

\begin{figure}[t]
\centering
\includegraphics[width=100.mm]{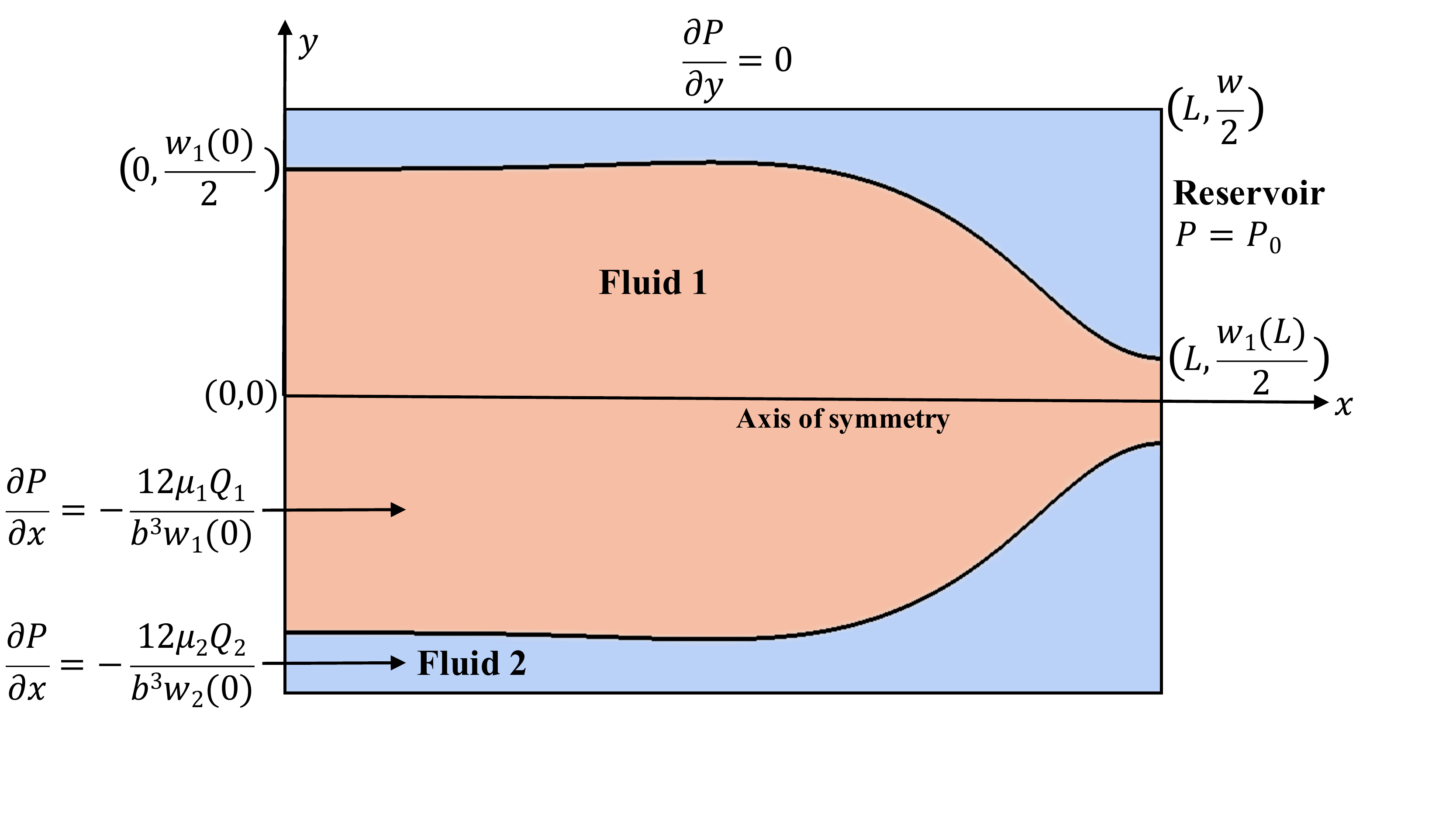}
\caption{Schematic of the flow domain and the corresponding boundary conditions for the Hele--Shaw cell
of width $w$.}
\label{fig:fig1}         
\end{figure}
We adopt the classical steady Hele--Shaw flow equations,
which is a simple description of the flow of a viscous Newtonian liquid
between two horizontal plates separated by a thin gap, with surface tension
\begin{equation}
{\ensuremath{\mathbf{u}(x,y,t)}} = \frac{[b(x,y)]^2}{12 \mu(f(x,y,t))} \left\{- \nabla P(x,y,t) + 
\gamma \kappa \delta_S \hat{{\ensuremath{\mathbf{n}}}}\right\},
\label{eq:HS}
\end{equation}
where $\mu(f(x,y,t))$ is the viscosity,
$b(x,y)$ the depth of the Hele--Shaw cell, $P(x,y,t)$ the local pressure, 
$\delta_S$ the interface Dirac delta function, $\kappa$ the in-plane curvature,
$\gamma$ the interfacial surface tension, 
and $\hat{{\ensuremath{\mathbf{n}}}}$ the unit normal. $f(x,y,t)$ is 
the volume of fluid function defined as $1$  \mbox{inside fluid 1} and $0$
\mbox{inside fluid 2}, for a two fluid system (see Fig.~\ref{fig:fig1}). 
The evolution of $f(x,y,t)$ satisfies the
advection equation
\begin{equation}
\partial_t f(x,y,t) + \nabla \cdot ({\ensuremath{\mathbf{u}(x,y,t)}} f(x,y,t))= 0.
\end{equation}
The numerical discretization and 
validations are described in detail in (Afkhami and Renardy 2013).
This simplified model can significantly reduce the computational cost and complexity, 
while allowing to keep the essential features of the flow. 
Furthermore, our computational model relies on the assumption that the 
pressure at the boundary,
where the fluid exits into the reservoir, is the uniform pressure in the reservoir.
This hypothesis is computationally investigated by comparing the numerical results
with the experimental measurements. In particular, 
we conjecture that the pressure boundary condition 
becomes important when the width of the neck is comparable to the depth of the microchannel.
A correction to the pressure boundary condition at the outflow is then proposed, relying only on
a single free parameter, which remains to be determined by comparison to experimental data.

\section{Results and Discussion}
We present the experimental observations and the numerical results based 
on the Hele--Shaw approximation combined with
a volume of fluid method for computing the interface motion and for modeling
the surface tension. 
Figure \ref{fig:fig1} shows a schematic of the flow domain and the corresponding 
boundary conditions used in the numerical model. 
For all the results here, we fix the viscosity ratio to 1 and 
set the same pressure gradient for each phase at the inlet ($x=0$). 
We also fix $b(x,y)=b$.
Experimental results in (Priest \etal 2006) and (Hein \etal 2015b) show 
that the transition of jet- to step-breakup regime is independent of the 
viscosity ratio when the results are presented in terms of the rescaled filament width $w_1(0)/b$
as a function of Ca. We therefore do not consider unmatched viscosities in this work. 
%and the channel aspect ratio $w/b=10$ or $30$. 
To model the reservoir, we set a zero pressure 
($P_0$ is the reference pressure in the reservoir) at the outlet ($x=L$). 
%We carry out direct simulation when varying the outflow pressure boundary condition
%to investigate the effects of the applied boundary condition on jet width across the domain.
We account for the influence of the growing three-dimensional droplet in the reservoir 
and the reservoir itself with a simple pressure boundary
condition at where the fluid exits into the reservoir. 
Specifically,
the outflow pressure boundary condition is defined as
\begin{equation}
P(L,y,t) = \left\{
  \begin{array}{ll}
    P_0 + A(2\gamma/b) & \quad \mbox{$-w_{1}(L)/2\le y \le w_{1}(L)/2$}\\
    P_0 & \quad \mbox{otherwise}
    \end{array} \right.
\label{eq1} 
\end{equation}
where $A$ is the pressure correction parameter and the out-of-plane curvature is assumed to be constant
and equal to $2/b$, i.e.~$2\gamma/b$ represents the Laplace pressure at the neck ($x=L$). 
Next, we examine this simple pressure boundary condition in both the jet-breakup regime,
where $w_1(L) \gg b$, and
close to the transition to step-breakup regime, where $w_1(L) \sim b$.
We will show that the pressure correction is
needed, when close to the transition to step-breakup regime, for reasonable agreement 
with physical experiments. We however note that 
since the pressure beyond the step is unknown, we can only speculate
either the inflating three-dimensional drop in the reservoir or
the out-of-plane curvature to be the physical origin of the pressure correction.
We leave a thorough study of these effects, including fully three-dimensional computations, for future work. 

\begin{figure}[t]
\centering
\includegraphics[width=90mm]{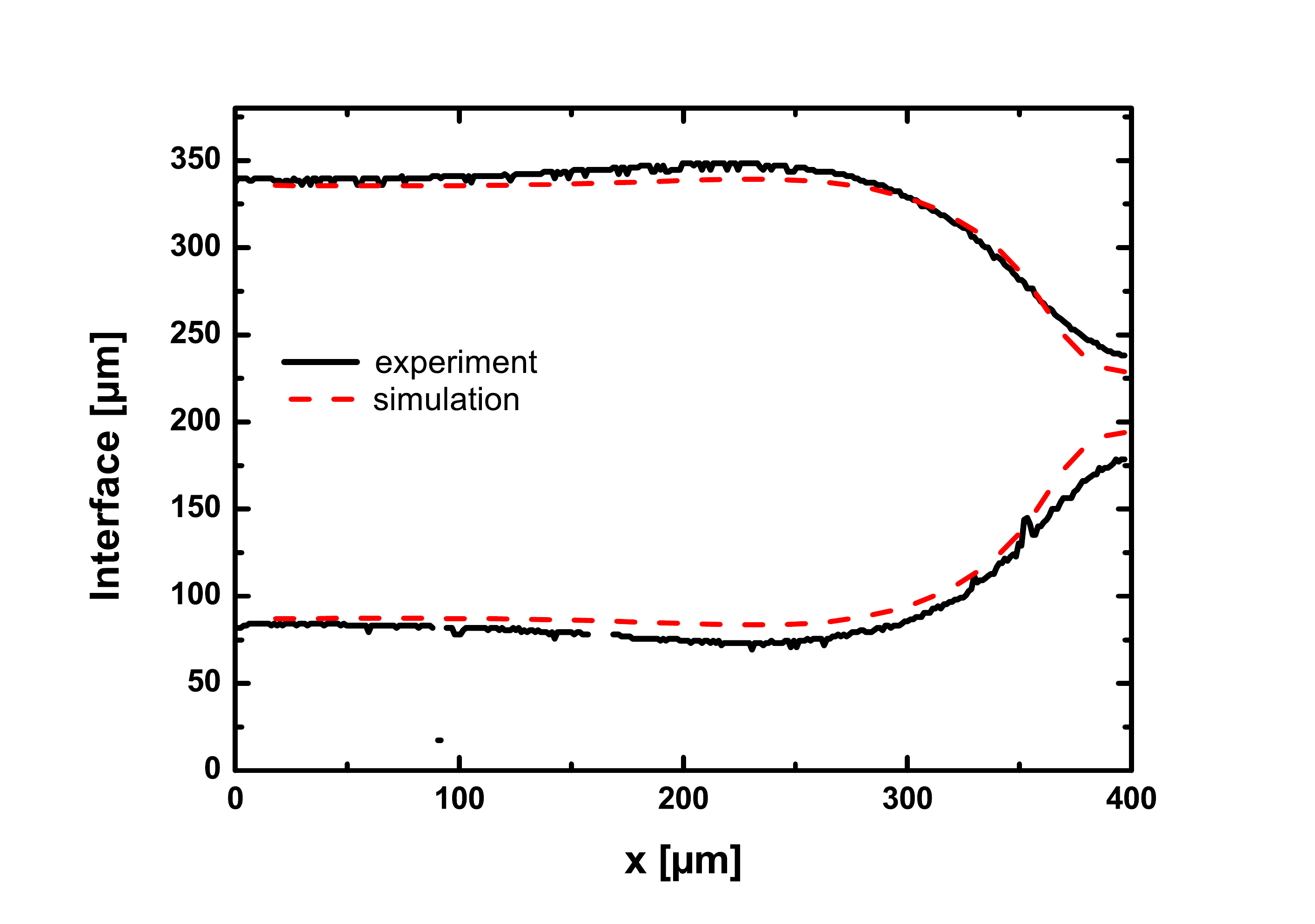}
\caption{Comparison of the numerical results with the experiment for Ca $= b^2\left[dp_1/dx\right]_{x=0}/(12\gamma)=0.019$ and $w_{1}(0)/b=12.76$, when $A=1$. }
\label{fig:fig2}         
\end{figure}
Figure \ref{fig:fig2} shows the computationally predicted interfacial profile, 
in the (single) jet-breakup regime, in comparison with the experimental
measurement for $A=1$, i.e.~assuming no correction to the pressure boundary
condition at the exit. As shown, a very good agreement with experimental visualization is obtained.
Next, we will consider the breakup regime close to the transition from jet- to the step-breakup.
We are going to define the transition of the jet-breakup to the step-breakup
as when the dispersed phase forms a circular jet upon exiting into the reservoir.
In reality, this should happen when $w_1(L)\approx b$, because the dispersed phase will become
unstable below this width due to the Rayleigh-Plateau type of instability.
Thus, in practice, $w_1(L)\approx b$ is the smallest achievable width of the focusing neck.
In the following, we will discuss in more detail the effect of the pressure correction, $A$, on the focusing 
close to the transition from the jet- to step-breakup regime.  

\begin{figure}
\begin{center}
\begin{tabular}{cc}
\includegraphics[width=75mm,trim=0 83mm 0 0]{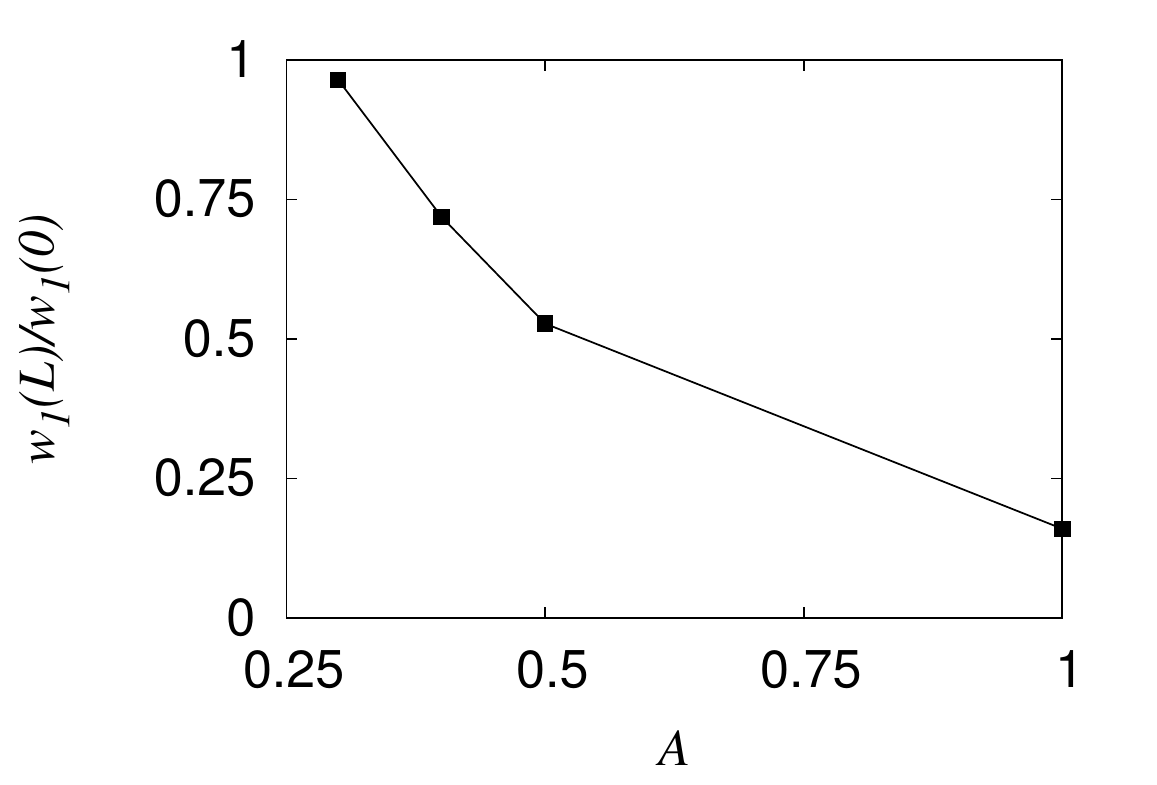}&
\includegraphics[width=55mm,angle= -90]{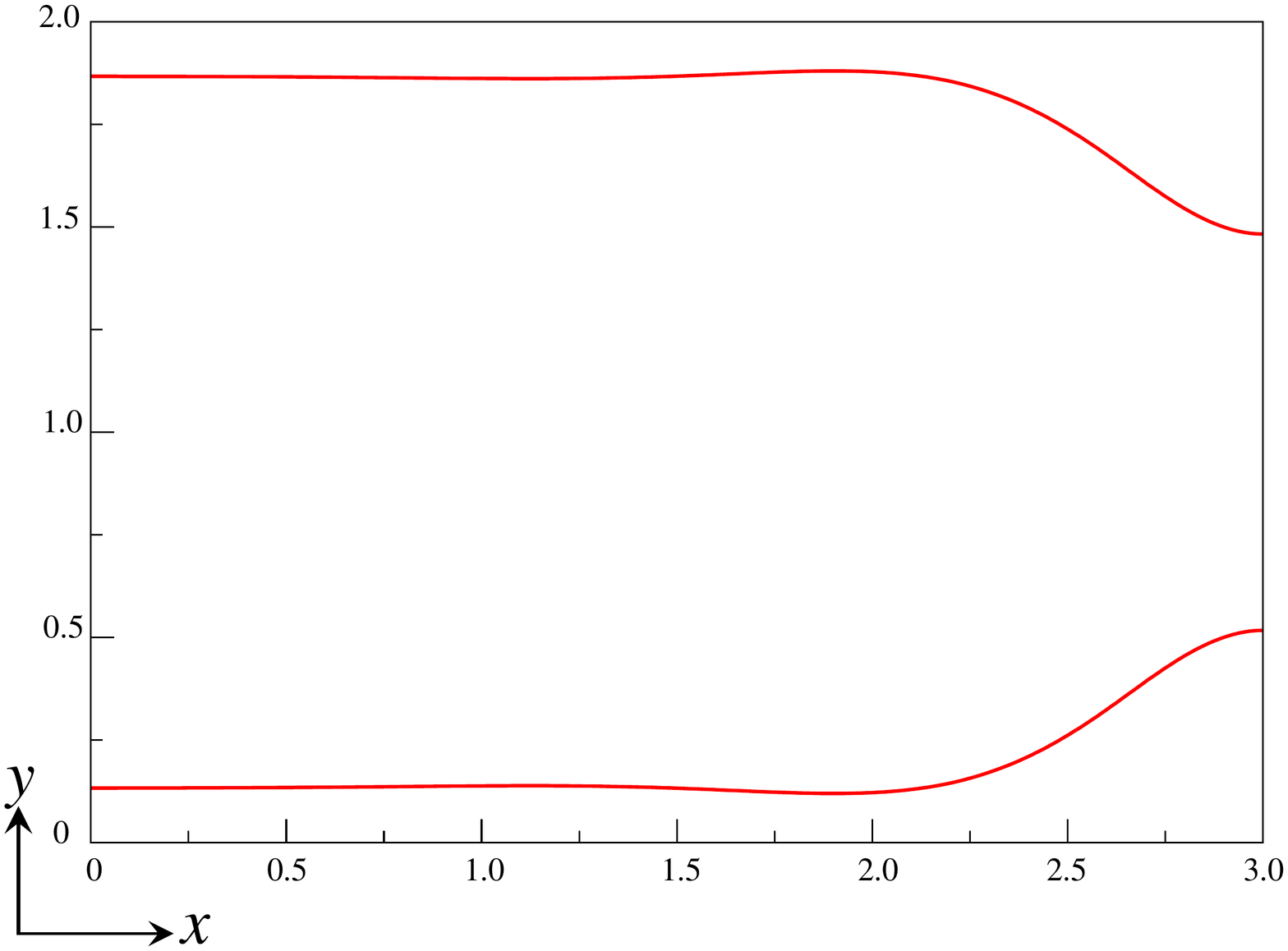}\\
(a)&(b)\\
\includegraphics[width=55mm,angle= -90]{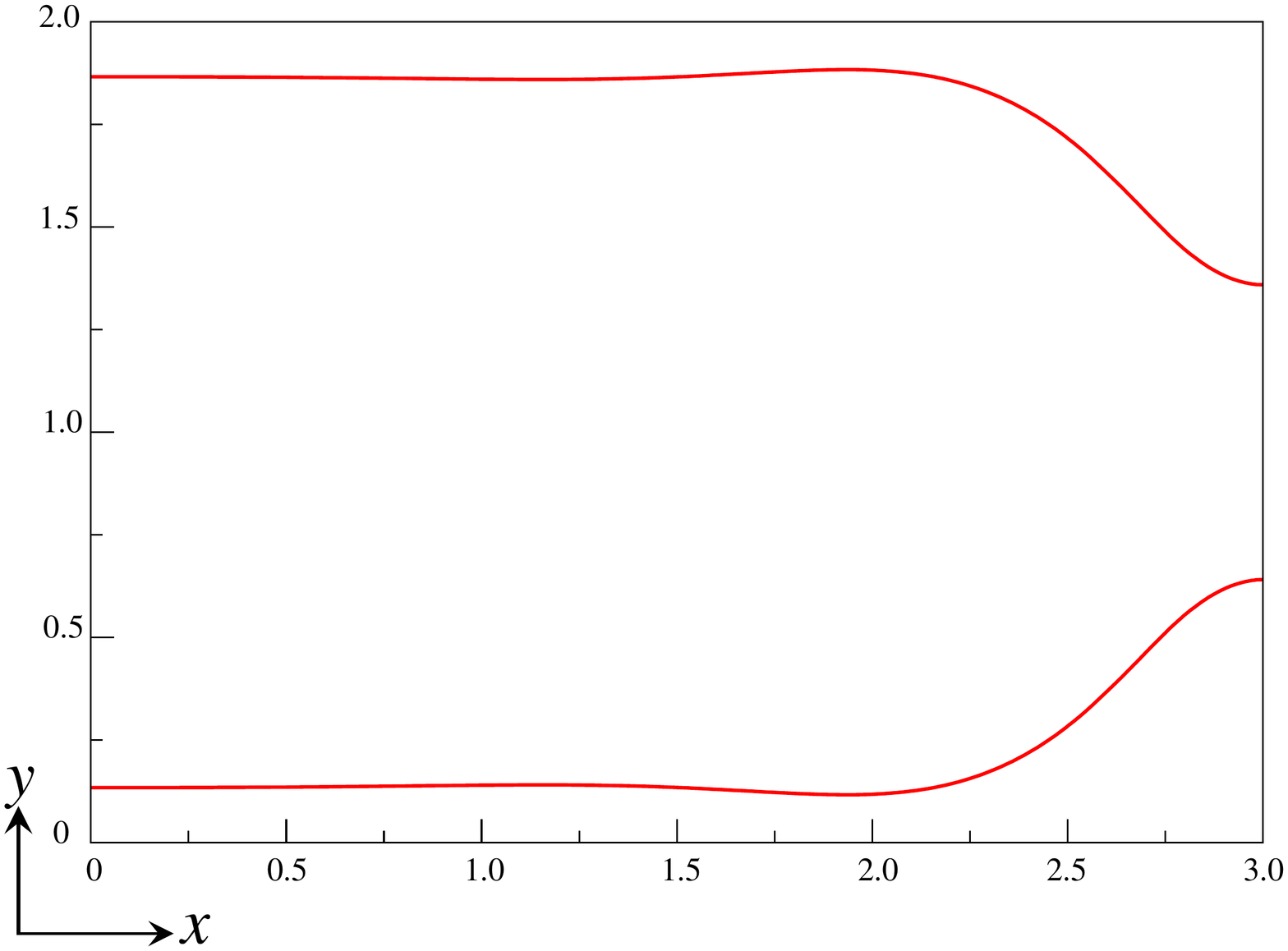}&
\includegraphics[width=55mm, angle= -90]{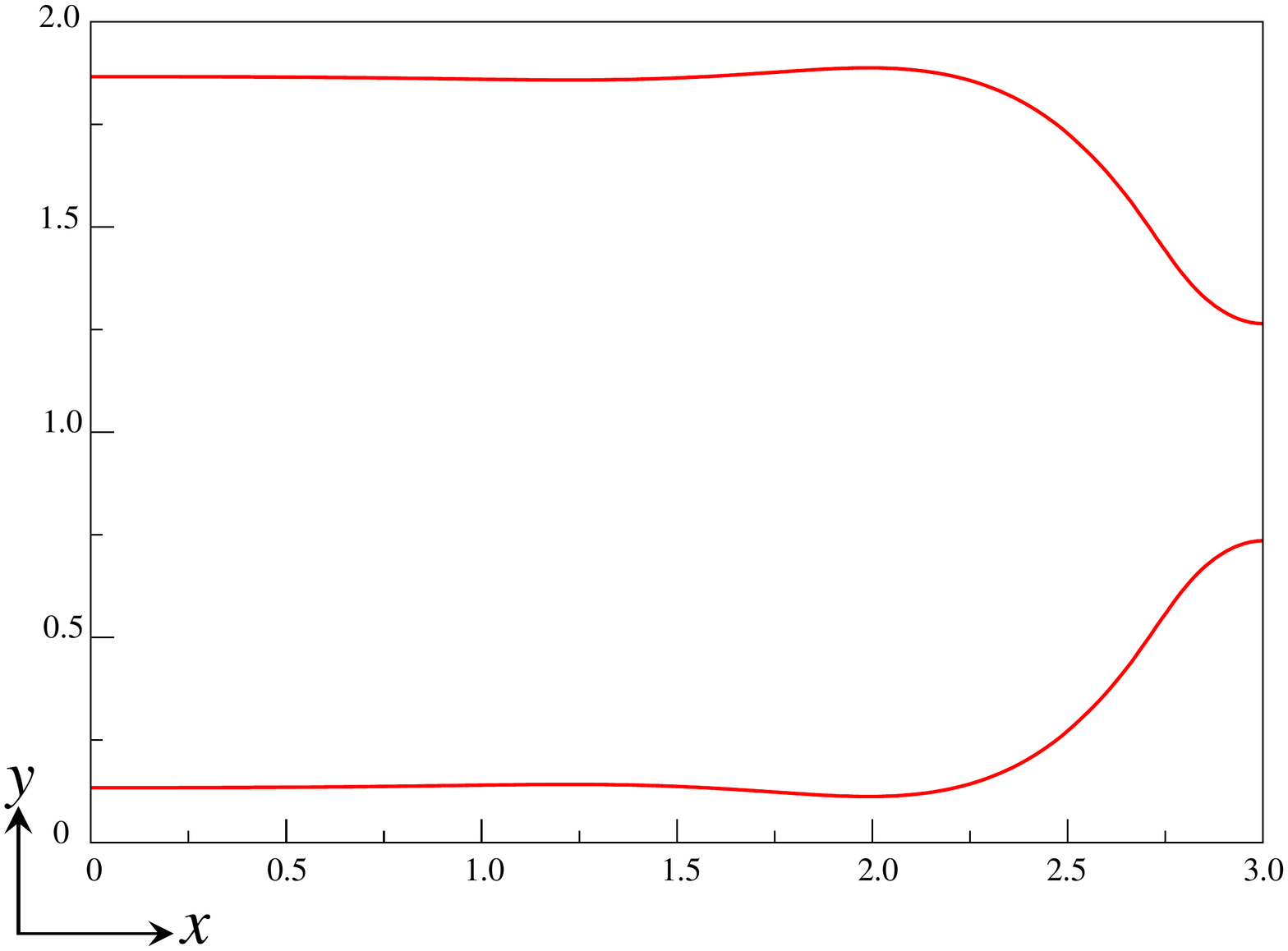}\\
\\
(c)&(d)
\end{tabular}
\end{center}
\caption{(a) $w_{1}(L)/w_{1}(0)$ as a function of $A$.
The shape of the capillary focusing with (b) $A=0.3$, (c) $0.4$, and (d) $0.5$.
The neck widens as $A$ is decreased. For these simulations Ca $=0.01$ and $w_{1}(0)/b=10$.} 
\label{fig3}
\end{figure}
Figure \ref{fig3}(a) presents the width of the focusing neck at the exit, 
$w_{1}(L)$, normalized by width of the dispersed phase at the inlet, $w_{1}(0)$, 
as a function of the pressure correction, $A$.
%We choose $A=0.3,0.4,0.5$ and $1$.
The results show that decreasing the pressure in the dispersed
phase at the step by a factor $3$ results in an increase in
$w_{1}(L)$ by a factor of $6$.
%Our experiments show that the instability should occur at Ca $\approx 0.01$ when 
%$w_{1}(L)/w_{1}(0)\approx 0.71$ corresponding to $A\approx 0.4$. 
We therefore show that the pressure boundary condition can play 
an important role when the width of the focusing neck becomes 
comparable to the depth of the channel. 
Figures \ref{fig3}(b-d) show the shape of the capillary focusing for three different 
pressure correction parameters used in Fig.~\ref{fig3}(a). 
As demonstrated, the capillary focusing becomes less
pronounced with a decrease of the pressure in the dispersed phase at the exit.
We note again that when $A=1$, the pressure inside the dispersed phase at the 
exit boundary is just $2\gamma/b$, i.e.~the Laplace pressure of a meniscus of mean curvature $2/b$.

As in (Hein \etal 2015a), we define the critical Ca number as the capillary number 
for which the jet-breakup transitions to a step-breakup for a fixed $w_{1}(0)/b$.
As discussed above, and shown experimentally in (Hein \etal 2015a), this transition
occurs when $w_1(L) \approx b$. For $w_{1}(0)/b =10$, the experiments of  (Hein \etal 2015a)
finds the critical Ca $\lesssim 0.005$. 
We next investigate whether we can find an $A$ 
that permits $w_1(L)\approx b$ for $w_{1}(0)/b=10$
at Ca close to the experimentally observed transition from jet- to step-breakup regime.
For each $A$, we vary the Ca number to obtain the
value of Ca for which $w_1(L)\approx b$. 
We find that only for $A=0.4$, we can obtain $w_1(L)\approx b$
for Ca $= 0.004$. From this analysis, we can infer that the transition from
jet- to step-breakup is reached at Ca $= 0.004$ 
only if the corrected pressure boundary condition at the outflow is employed.
This corrected pressure boundary condition corresponds to a significantly smaller 
Laplace pressure inside the dispersed phase at the exit, which could be the consequence of a 
smaller out-of-plane curvature than considered when $A=1$; i.e. $2/b$.
One possible physical explanation can be that in the step-breakup regime, upon the 
formation, the droplet is initially confined and rapidly `feels' the reservoir, causing it to 
assume a smaller out-of-plane curvature than $2/b$.  
We therefore conjecture that the capillary focusing is strongly sensitive to the values
of the considered out-of-plane curvatures at the exit, and therefore the 
discrepancy between the numerical results and the experimental data is most likely
explained through this sensitivity. We thus show that our correction to the outflow pressure
boundary condition can be an effective way to achieve numerical results that agree reasonably well
with experimental measurements, without resorting to fully three-dimensional simulations.

\begin{figure}
\begin{center}
\includegraphics[width=70mm, angle= -90]{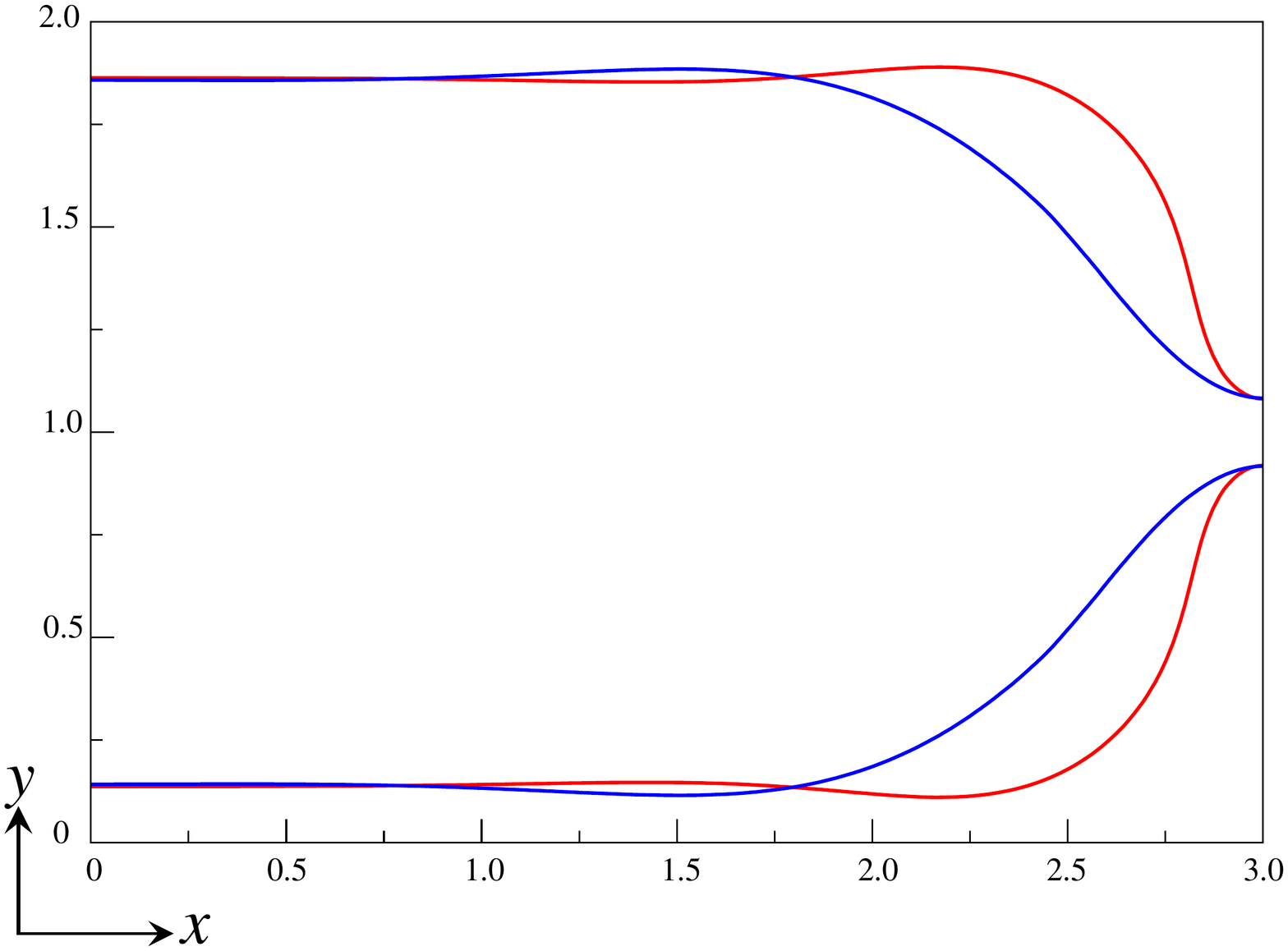}\\
(a)\\
\includegraphics[width=87mm]{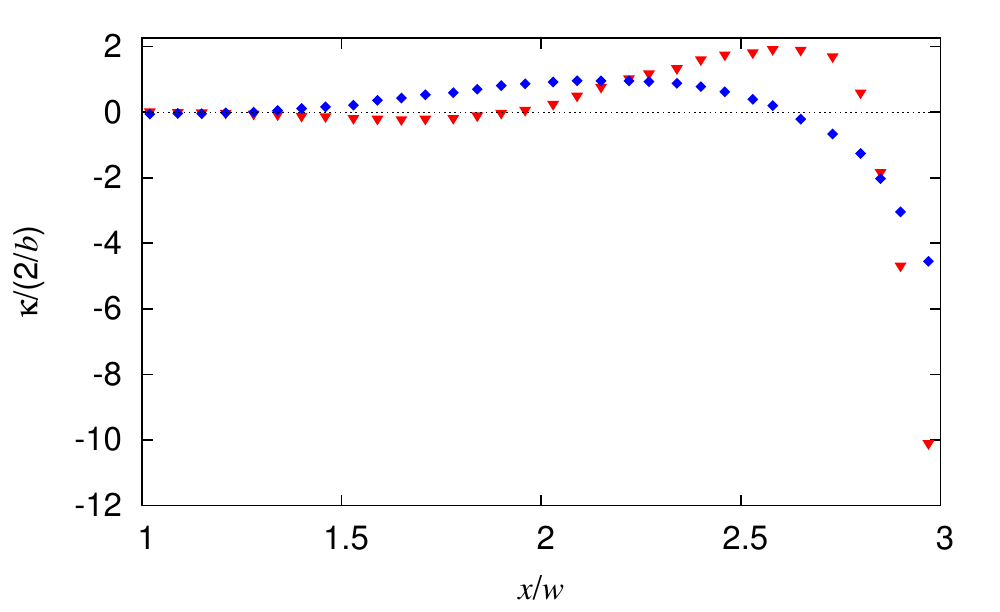}\\
(b)
\end{center}
\caption{(a) The shape of the capillary focusing for Ca $=0.01$ 
with no pressure boundary condition correction at the outflow, 
$A=1$ (red), and for Ca $=0.004$ with a pressure boundary 
condition correction at the outflow, $A=0.4$ (blue); $w_1(L)\approx b$.
(b) Computed interface curvature normalized by $2/b$, the 
magnitude of the curvature at the outflow ($x=L$), as a function
of $x/w$ for Ca $=0.01$ and $A=1$ ({\color{red}$\blacktriangledown$}), 
and for Ca $=0.004$ and $A=0.4$ ({\color{blue}$\blacklozenge$});
the dashed line, $\kappa=0$, is plotted for visual assistance for when the curvature changes sign.
$w_{1}(0)/b=10$.} 
\label{fig4}
\end{figure}
However, the precise prediction of the critical Ca number (i.e.~the Ca for which $w_1(L)\approx b$), 
by introducing the pressure correction $A$,
comes with the expense of a reduced focusing effect. 
Figure \ref{fig4}(a) shows that we obtain $w_1(L)\approx b$ for Ca $=0.01$ when $A=1$
and for Ca $=0.004$ when $A=0.4$.
We remind that in (Hein \etal 2015a), 
the breakup regime transition is reported to occur at Ca$\lesssim 0.005$.
Figure \ref{fig4}(a) shows that the focusing effect is much stronger when
$A=1$ and Ca $=0.01$ than when $A=0.4$ and Ca $=0.004$. This is somehow counterintuitive.
It shows that if we do not consider any pressure correction at the outflow, i.e.~$A=1$, 
a much larger Ca must be attained in order to obtain $w_1(L)\approx b$, leading
to a more profound capillary focusing effect.
%, while with a pressure correction, i.e.~$A=0.4$, the capillary focusing is more gentle.
%In other words, we cannot match the interface shapes when we use $A=1$ and $0.4$
%despite the fact that in both cases, $w_1(L)\approx b$. 
%We also numerically show that the dispersed phase bulges into the continuous phase before it
%thins out at the exit, in accordance with the experimental observations \cite{}. 
Figure \ref{fig4}(b) demonstrates the local change of the interface shape in the focusing region
by plotting the computed interface curvature for Ca $=0.01$ and $A=1$, and for Ca $=0.004$ and $A=0.4$.
The results show the abrupt change in the curvature in the narrowing region; curvature becomes negative 
in the focusing region. 
As shown, the increase in the capillary number results in the decrease of the length scale over which
the curvature changes sign, i.e.~the focusing region length scale.
We also note that with $A=1$, we arrive at $\kappa=-2/b$ at the outlet ($x=L$),
while smaller $A$ results in a smaller Laplace pressure at the outlet.

\subsection{Transitions of breakup regimes}
\begin{figure}
\begin{center}
\begin{tabular}{cc}
\includegraphics[width=0.4\textwidth, trim=0 11mm 155mm 0, clip=true]{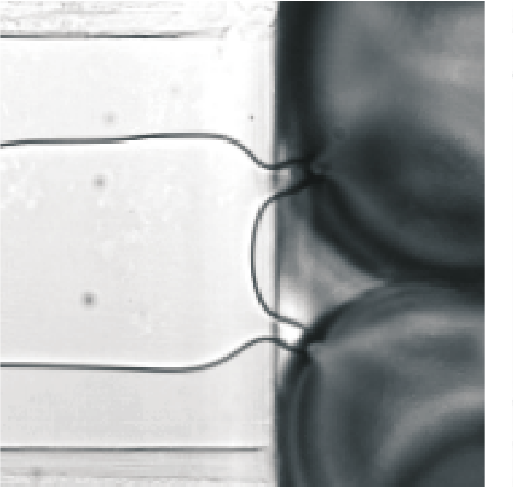}&
\includegraphics[width=0.4\textwidth, trim=155mm 11mm 0mm 0, clip=true]{fig7.pdf}\\
(a)&(b)\\
\includegraphics[width=0.5\textwidth]{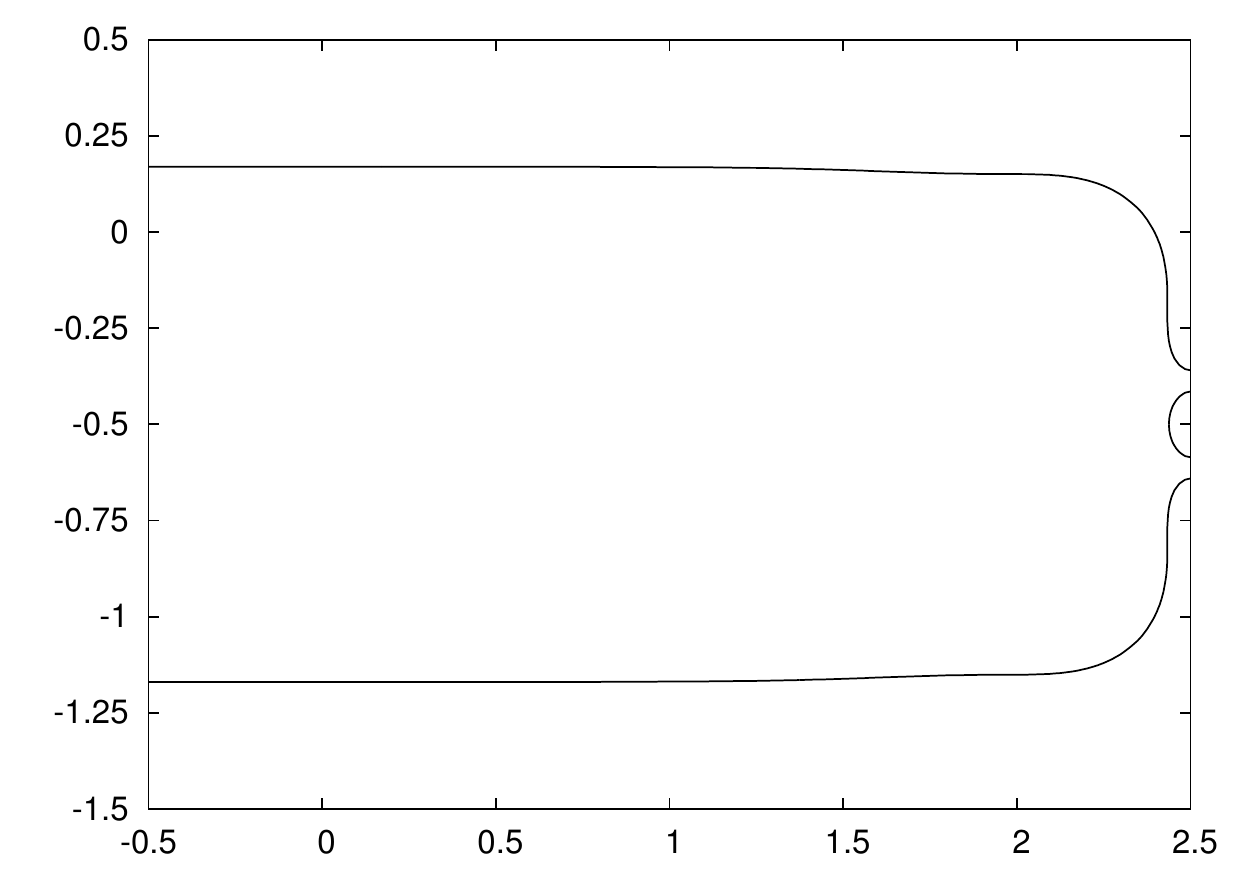}&
\includegraphics[width=0.5\textwidth]{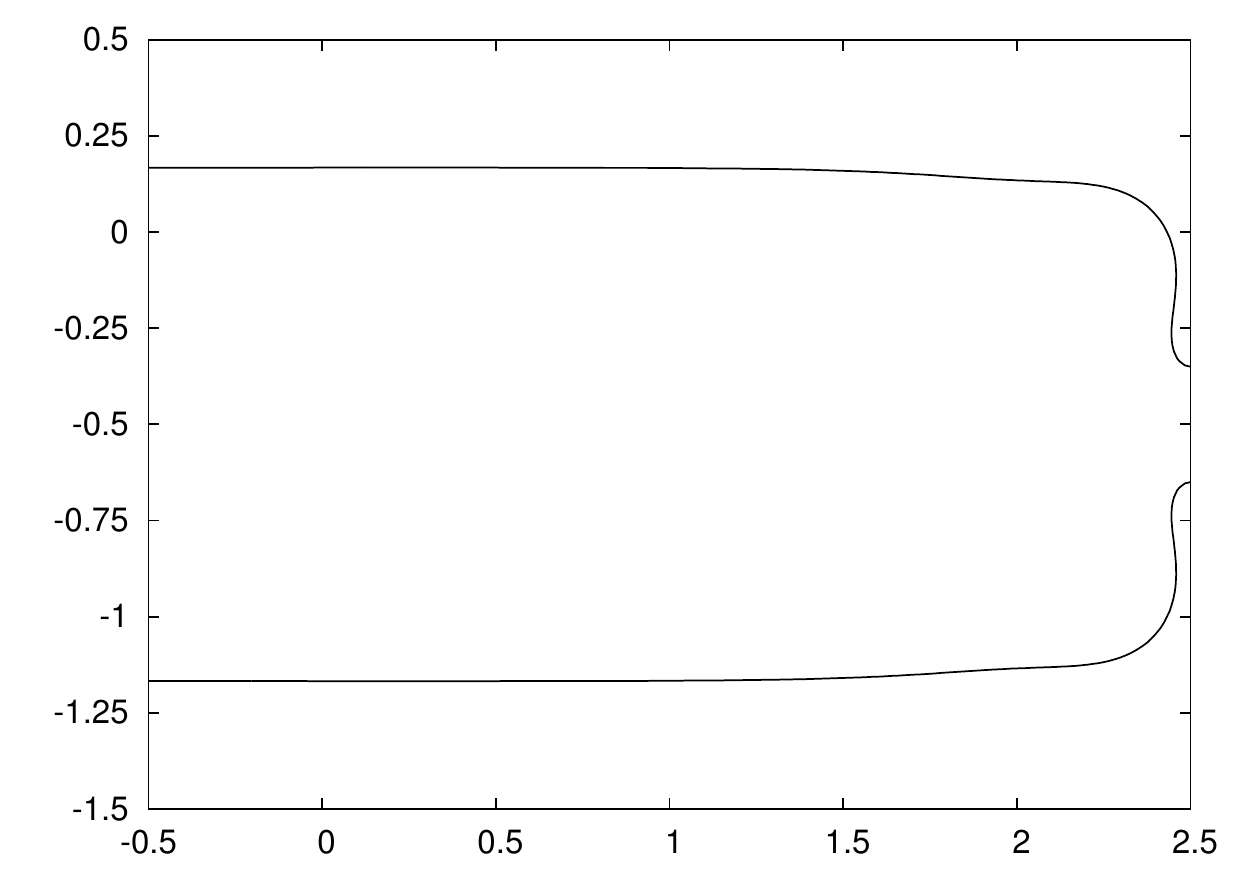}\\
(c)&(d)
\end{tabular}
\end{center}
\caption{ 
It is shown that a double jet-breakup can occur depending on 
$w_1(0)/ b$ and the Ca number. For a fixed $w_1(0)/ b$,
increasing Ca number results in collapsing of double jets 
into a single jet.
Optical micrographs showing two regimes of droplet breakup;
(a) a double jet-breakup for Ca $=0.0223$ and $w_{1}(0)/b=16.4$ and 
(b) a single jet-breakup for Ca $=0.039$ and $w_{1}(0)/b=19.6$.
Simulations showing (c) a double jet-breakup for Ca $=0.0125$ and 
(d) a single jet-breakup for Ca $=0.025$; $w_{1}(0)/b=20$.
For these results $w/b=30$.}
\label{fig5}
\end{figure}
In (Hein \etal 2015b), multiple breakup regimes are addressed. 
In their work, the authors show that,  when $w_1(0)/ b\gtrsim 12$,
a variety of breakup regimes can be observed 
depending on the Ca number. 
For example, they show that when $w_1(0)/ b$ is fixed and $w_1(0)/ b \gtrsim 15$,
a double jet-breakup can occur, for a particular range of small Ca numbers, which 
collapses into a single jet-breakup by increasing the Ca number. 
In Fig.~\ref{fig5}(a-b), we show the experimental observation of these 
two breakup scenarios. We attribute the double jet-breakup
to the strong confinement. The transition from double 
to single jet-breakup is discussed in details in (Hein \etal 2015b).
Here we also numerically show the existence of these breakup regimes.
Figures \ref{fig5}(c-d) show simulation results of different breakup regimes 
for Ca$=0.0125$ (double jet-breakup) and Ca=$0.025$ (single jet-breakup) 
when $w_{1}(0)/b=20$. This transition is in agreement with our experiments 
previously documented in (Hein \etal 2015b). 
For these simulations, we use $A=1$. These unprecedented results
provide excellent insight into the capillary self-focusing phenomena, 
capturing multiple breakup regimes using our simple pressure boundary condition. 
The numerical results shown in Figs. \ref{fig5}(c-d) also exhibit
the strong capillary focusing effect when increasing confinement.
Figure \ref{fig5}(d) shows that the front of the filament is almost flat 
at the exit, in agreement with the profile in Fig.~\ref{fig5}(b). Upon decreasing the Ca number,
i.e.~increasing the capillary focusing effect,
such a wide and flat interface allows easy formation of another jet at the exit 
(Fig. \ref{fig5}(c)).
Our experiments show that for a double jet-breakup, 
droplets that are formed in the reservoir (see Fig. \ref{fig5}(a))
grow and elongate while traveling downstream;
during this elongation, the two tips formed at the step are pushed outwards. 
The absence of this effect explains the difference in the location of the 
two tips which are more centered in the simulation.

\section{Conclusions}
We confirm the validity of our numerical simulations, based on the classical Hele--Shaw
formulation and a  simple pressure boundary
condition at which the fluid exits into the reservoir, by comparing with experimental results.
Despite the inherent assumptions, the computational framework
describes the capillary self-focusing consistently with our experimental findings.
For the jet-breakup regime and sufficiently large Ca numbers, 
we show that the numerical model well captures the focusing profile without any
correction to the pressure boundary condition. 
Close to the transition from a jet-breakup
regime to a step-breakup regime, a pressure correction is required in order to predict 
the transition threshold. The correction is most likely due to the effects of
the expanding three-dimensional droplet in the reservoir
or an out-of-plane curvature that becomes smaller than $2/b$ 
because of the sudden relief of the confinement at the step,
or a combination of these.
For further consideration of these effects, fully three-dimensional computations 
are needed. We leave this extension for future work. 

We also conjecture that the pressure boundary condition 
becomes important when the width of the focusing region 
is comparable to the depth of the shallow microchannel.
The corrected pressure boundary condition at the outflow takes into account
a single free parameter, to be determined by comparison with experimental data.
Our numerical results also provide an excellent insight into the capillary 
self-focusing phenomena when there exists a strong confinement. In particular,
consistently with our experimental results, we show that our numerical simulations 
can identify a double jet-breakup regime and its transition to a single jet-breakup regime
upon decreasing the confinement or increasing the Ca number.
Ongoing work in microfluidics drop generation makes our results
highly relevant for technical applications. For example,
for droplet based microfluidic systems (see e.g.~Seemann \etal 2012),
when using channel geometries such as the ones in this work,
the occurrence of multiple breakup sites at a topographic step
on the same filament
will be likely or even desired, e.g.~for the simultaneous particle 
or cell encapsulation or filtering, providing a novel tool in the droplet microfluidics applications.

%Ongoing work in microfluidics drop generation makes our results
%highly relevant for technical applications.
%We therefore conclude that imposing a constant pressure boundary condition at the outflow
%with a correction is not sufficient and that the slope at the outlet should also be considered.
%This slope is not fixed and perhaps vary with time as the breakup occurs. 
%A more accurate boundary condition therefore should consider the emerging droplet in the reservoir.  

%\section*{Acknowledgments}
\ack
This work was partially supported by the grant Nos.~DFG-GRK1276 and NSF-DMS-1320037.
Authors gratefully acknowledge Dr.~Jean-Baptiste Fleury (Saarland University) for helpful discussions.

\section*{References}
\begin{harvard}
\item[] Priest C, Herminghaus S, and Seemann R 2006 Generation of monodisperse gel emulsions in a microfluidic device {\it Appl.~Phys.~Lett.} {\bf 88} 024106
\item[] Malloggi F, Pannacci N, Attia R., Monti F, Mary P, Willaime H,
Tabeling P, Cabane B, and Poncet P 2010 Monodisperse Colloids Synthesized with Nanofluidic Technology {\it Langmuir} {\bf 26} 2369-2373
\item[] Afkhami S and Renardy Y 2013 A volume-of-fluid formulation 
for the study of co-flowing fluids governed by the Hele--Shaw equations {\it Phys.~Fluids} {\bf 25}
082001
\item[] Hein M, Afkhami S, Seemann R, and Kondic L 2015a Capillary focusing close to a topographic step: shape and instability of confined liquid filaments {\it Microfluid.~Nanofluid.} {\bf 18} 911-917
\item[] Li Z, Leshansky A, Pismen L, and Tabeling P 2015 Step-emulsification in a microfluidic device {\it Lab Chip} {\bf 15} 1023-1031
\item[] Hein M, Fleury J-B, and Seemann R 2015b Coexistence of different droplet generating instabilities: new breakup regimes of a liquid filament {\it Soft Matter} {\bf 11} 5246-5252
\item[] Seemann R, Brinkmann M, Pfohl T, and Herminghaus S 2012 Droplet based microfluidics
{\it Rep.~Prog.~Phys.} {\bf 75} 016601.
\end{harvard}

\end{document}